\documentstyle[11pt,epsfig]{article}
\begin{document}
\def\boxit#1{\vcenter{\hrule\hbox{\vrule\kern8pt
      \vbox{\kern8pt#1\kern8pt}\kern8pt\vrule}\hrule}}
\def\Boxed#1{\boxit{\hbox{$\displaystyle{#1}$}}} 
\def\sqr#1#2{{\vcenter{\vbox{\hrule height.#2pt
        \hbox{\vrule width.#2pt height#1pt \kern#1pt
          \vrule width.#2pt}
        \hrule height.#2pt}}}}
\def\square{\mathchoice\sqr34\sqr34\sqr{2.1}3\sqr{1.5}3}
\def\Square{\mathchoice\sqr67\sqr67\sqr{5.1}3\sqr{1.5}3}
\def\AJP{{\it Am. J. Phys.}}
\def\AM{{\it Ann. Math.}}
\def\AP{{\it Ann. Phys.}}
\def\CQG{{\it Class. Quantum Grav.}}
\def\GRG{{\it Gen. Rel. Grav.}}
\def\JMP{{\it J. Math. Phys.}}
\def\JP{{\it J. Phys.}}
\def\JSIRAN{{\it J. Sci. I. R. Iran}}
\def\NC{{\it Nuovo Cim.}}
\def\NP{{\it Nucl. Phys.}}
\def\PL{{\it Phys. Lett.}}
\def\PR{{\it Phys. Rev.}}
\def\PRL{{\it Phys. Rev. Lett.}}
\def\PRp{{\it Phys. Rep.}}
\def\RMP{{\it Rev. Mod. Phys.}}
\title{\bf  Intrinsic Gravitomagnetism and Non-commutative effects}
\author{{\small Behrooz Malekolkalami,}\footnote{b.malakolkalami@uok.ac.ir}\ \ {\small\
         Awat Lotfi}\footnote{alwf12.kurd@gmail.com}\\
        {\small Department of Physics,  University of Kurdistan, Pasdaran St., }\\
        {\small Sanandaj, Iran}}
\date{\small \today}
\maketitle
\begin{abstract}
The two equations of motion for a test particle are compared with each other. One is in the Non--commutative space involving a static rigid sphere (as a source of central force) and the other is in the usual space involving a slowly stationary rotating sphere. The comparison tells us that,  the effects of  Non--commutativity  is analogous to gravitational effects of rotation of the sphere. That is, non--commutativity imitates the effects of the intrinsic gravito--magnetic field of the sphere.
\end{abstract}

\bigskip

\section{Introduction}
\indent
There are some profound analogies and great differences between gravitation and
classical electrodynamics. Many of these differences are due to the tensorial nature of
the gravitational interaction and to the nonlinearity of the Einstein field equations.
The effect of mass current  in general relativity leads to some unexpected
and profound analogies with the magnetic interactions. These effects have been
expressed in the study of Gravito--Magnetism. There are two types of mass currents in gravity:

The first type  is produced by translational motion of matter.
It generates an extrinsic Gravito--Magnetic (\textbf{GM}) field which depends on the frame of
reference of observer and hence it can be  eliminated in the rest frame
of the matter.

The second type of mass current is caused by
the intrinsic rotation of matter around the body's center of mass. Intrinsic
GM field is directly related to the proper body's angular momentum (spin) and
most of researches  in gravito--magnetism have focused on the discussion of its various
properties. To  give a comprehensive review of various aspects of the intrinsic gravito--magnetism and to study at greater depth, the reader may refer to  worthwhile textbooks as~\cite{ciuwheel95, Wheeler}.

One of the important aspects of intrinsic gravito--magnetism is related to its effects  in shaping the test particle orbit about a rotating central mass. Most important of these effects can be outlined as follows:

(i) the precession of a gyroscope
in orbit,

(ii) the precession of orbital planes
in which a mass orbiting a large
rotating body constitutes a gyroscopic system whose orbital axis will
precess,

(iii) the precession of the pericenter of the orbit of a test mass about a
massive rotating object.

The first and second cases  are due to the frame-dragging  effects and the \emph{Gravity Probe B} and the \emph{LAGEOS satellites} attempted to measure of these effects\cite{Probe}.
Here, we are concerned  about the third case (iii) in which the massive rotating object is a slowly rotating
sphere which gravitational (gravito-electric and magnetic) fields are well known, e. g. \cite{mash2000}. As we know, in absence of rotation, the trajectories of a test particle are Keplerian orbits, but rotation (which generates intrinsic GM effects) causes a precession motion. The effects of such precession has been discussed and illustrated in some of the works as\cite{Castaneda}.

On the other hand, the Non--Commutative (\textbf{NC}) geometry has played an increasingly important role, more notably, in the attempts to
understand the space–-time structure at very small distances. The NC geometry provides valuable tools for the study of physical theories in classical and quantum level. Due to the large volume of papers
related to this subject, we will avoid referring the papers here. Our interest here will be to focus on NC effects in central force orbits which has been studied in literature, e. g. \cite{Romero, Romero1, Mirza}.

Owing to the symplectic structure of the NC space, it is evident that the NC effects (which are described by the NC parameters  $\theta_{ij}$ and  $\beta_{ij}$)\footnote{The NC parameters  $\theta_{ij}$ and  $\beta_{ij}$ are related to the coordinate and momentum sectors of the phase space, respectively.} are inevitably presented in the motion equations. In other words, the NC parameters deform the motion equations (with respect to the usual space). Comparing these motion equations with those  come from the usual space, one finds the NC effects can interpreted as presence of additional (or new) force fields. A familiar example of these fields is the magnetic ones and indeed, under special conditions, by comparing the motion equations (in the NC and usual spaces), one can deduce that effect of the NC parameters on the motion equation is equivalent to presence of a constant
magnetic field in the usual space. The equivalencies of this type have been demonstrated  in the classical and in the quantum  perspectives, e. g. \cite{Djamai}.

Analogy between gravitation (in the weak field limit) and electromagnetism permits the extension of such equivalency between the NC effects and  GM field which has been shown in previous work \cite{BF}. More clearly, the authors show the equivalency between NC parameter $\beta_{ij}$  and  extrinsic (\emph{constant}) GM field. In the present work, it will be shown that, there is an equivalency between  NC parameter $\theta_{ij}$ and  intrinsic (\emph{variable}) GM field. We must note that the case of extrinsic (\emph{constant}) GM field (in the previous work) is somewhat speculative (\emph{gedanken}), that is, it may has not a real analogous in the nature or experimental evidences. While the case of intrinsic GM field has known real analogous in the nature, as rotating stars or  black holes and the other  celestial  massive objects.  The work is organized as follows:

In section 2, we introduce the  equations of motion for a test particle moving in the two space. The first is in a NC space which involves a central force field and second is in usual space involves a massive rotating  sphere. In section 3, the comparison of two sets of motion equations reveals  the similarity between the NC effects and the intrinsic GM field. A summary of the work is given in the last section.
\section{Equations of Motion}
In the two following subsections, we give equations of motion  in the NC frame and in a gravito--electromagnetic field produced by the slowly rotating sphere.
\subsection{NC Space}
Define a NC classical mechanics  equipped with symplectic structure
\begin{eqnarray}\label{E1}
\{ x_{i}, x_{j}\}= \theta_{ij},\quad \{ x_{i}, p_{j}\}= \delta_{ij},
\quad \{ p_{i}, p_{j}\}= 0, \label{eq:poisson}
\end{eqnarray}
where  the NC parameter
$\theta_{ij}$ is  real and anti--symmetric quantity\rlap.\footnote{The anti--symmetric property is usually defined in terms of Levi-Civita symbol $\epsilon_{ijk}$, that is $\theta_{ij}=\epsilon_{ijk}\theta_{k}$.}
Then, the Hamilton's equations corresponding  to NC structure (\ref{E1}) take the form
\begin{eqnarray}
&\dot x_{i}&=\frac{p_{i}}{m}+\theta_{ij}\frac{\partial V}{\partial x^{j}}, \label{eq:,}\\
&\dot p_{i}&=-\frac{\partial V}{\partial x^{i}},   \label{eq:h1}
\end{eqnarray}
which by elimination of momentum variable, become
\begin{equation}\label{E11}
m\ddot x_{i}=-\frac{\partial V}{\partial x_{i}}+
m\theta_{ij}\frac{\partial^{2 } V } { \partial x_{j}\partial x_{k}}
\dot x_{k }.
\end{equation}
Substituting the Keplerian  potential $V(r)=\frac{-k}{r}$ in (\ref{E11}), one gets
\begin{equation}\label{E2}
m\ddot x_{i}=-\frac{x_{i}}{r}\frac{k }{r^{2}}+m\epsilon_{ijk}\dot
x_{j} \Omega_{k}+m\epsilon_{ijk}x_{j}\dot
\Omega_{k}\label{eq:kepler}
\end{equation}
where
$$\Omega_{i}=\frac{k}{r^{3}}\theta_{i},$$
is similar to  an angular velocity.
Note that the second and third terms (in (\ref{E2})) are due to the NC effects which can be considered as representative of the new forces. For example,
The second term  is a Quasi--Coriolis force and thus it can be equivalent to a non-inertial force.

We restrict the non--commutativity to the $x-y$ plan,  which is equivalent to take
$\theta_1=\theta_2=0, \theta_3=\Theta$, or to take the angular velocity vector in the $z$ direction, namely
\begin{center}$\vec{\Omega}=(0, 0, \Omega= k \Theta/r^3)$,\end{center}
then the motion equations (\ref{E2}) in the spherical coordinates take the form
\begin{eqnarray}
m(\ddot r -r\dot \theta^{2}-r\dot \phi^{2}{\sin}^{2}\theta)=
-\frac{k}{r^2}+mr\Omega\dot \phi {\sin}^{2}\theta, \label{sphe1}\\
m \frac{d}{dt}(r^{2}\dot\theta)-mr^{2}\dot \phi^{2}{\sin}\theta{\cos}\theta=
-m r^{2}\Omega\dot \phi {\sin}\theta{\cos}\theta,\label{sphe2}\\
\frac{d}{dt}(mr^{2}\dot \phi {\sin}^{2}\theta)=mr{\sin}\theta
\frac{d}{dt}(r\Omega {\sin}\theta). \label{sphe3}
\end{eqnarray}
Without loss of generality, we take the motion in the plan   $\theta=\pi/2$, hence the latter equations are abbreviated as
\begin{eqnarray}
&m(\ddot r -r\dot \phi^{2})=&-\frac{k}{r^2}+mr\Omega\dot \phi=-\frac{k}{r^2}+km\Theta\frac{\dot \phi}{r^2}, \\
&\frac{d}{dt}(mr^{2}\dot \phi )=&mr\frac{d}{dt}(r\Omega)=-2mk\Theta\frac{\dot{r}}{r^2}, \label{eq:angular}
\end{eqnarray}
or
\begin{eqnarray}
&\ddot r=& r\dot \phi^{2}-\frac{k/m}{r^2}+k\Theta\frac{\dot \phi }{r^2} \label{eq:angular1}, \\
&\ddot \phi=&-2\frac{\dot r}{r}\dot \phi-2k\Theta\frac{ \dot{r}}{r^4}. \label{eq:angular}
\end{eqnarray}
\subsection{Slowly Rotating Sphere}
Here, we first introduce the Lagrangian for a test particle moving in a  stationary gravito-electromagnetic field and then it is employed to determine the motion equations of a test particle orbiting the rotating sphere.
Our Lagrangian is expressed in terms of the scalar and vector gravito-electromagnetic potentials $(\Phi, \textbf{A})$, that is\cite{mash2000}
\begin{equation}
{\cal L}= -mc^{2} ( 1 - \frac{v^2}{c^2})^{\frac{1}{2}} + m\gamma(1+ \frac{v^2}{c^2}) \Phi - \frac{2m}{c}\gamma {\bf v} \cdot {\bf
A},
\end{equation}
where $\gamma$ is the Lorentz factor. In the case of weak field and slow motion approximations, the above Lagrangian reads
\begin{equation}\label{D1}
{\cal L}= {\frac{1}{2}}mv^2 + m\Phi - \frac{2m}{c}{\bf v} \cdot {\bf A},
\end{equation}
which for  rotating sphere of mass $M$  and angular momentum $L=I\omega$ \rlap,\footnote{The angular velocity $\omega$ is to be taken in the $z$ direction and the moment of inertia about this direction is $I=2Ma^2/5$.} the potentials become
\begin{equation}\label{D2}
\Phi=-\frac{GM}{r}, \hspace{1.5 cm} {\bf A}=-\frac{GI\omega}{2c^2}\frac{\sin\theta}{r^2}\hat{\phi}.
\end{equation}
It should be noted that, the above potentials are  in the large distance approximation, namely
\begin{equation}\label{Sch}
r\gg2GM/c^2.
\end{equation}
Substituting potentials (\ref{D2})  in (\ref{D1}), gives\footnote{This Lagrangian can also be obtained by electrodynamics analogy\cite{Franklin}.}
\begin{equation}
{\cal L}= {\frac{1}{2}}m(\dot r^2+r^2\dot \theta^2+r^2\dot\phi^2\sin^2\theta) + \frac{GmM}{r} - \frac{GmI\omega}{2c^2}\frac{{\sin}^{2}\theta}{r}\dot \phi,
\end{equation}
and hence, the Lagrange equations become
\begin{eqnarray}
\ddot r =r\dot \theta^{2}+r\dot \phi^{2}{\sin}^{2}\theta
-\frac{GM}{r^2}+\frac{GI\omega}{2c^2}\frac{{\sin}^{2}\theta}{r^2}\dot \phi, \label{sphe1}\\
\ddot \theta= -2\frac{\dot{r}}{r}\dot\theta+\dot \phi^{2}{\sin}\theta{\cos}\theta
-\frac{GI\omega}{2c^2} \frac{{\sin}2\theta}{r^3} \dot \phi,\label{sphe2}\\
\ddot \phi=-2\frac{\dot{r}}{r}\dot\phi-2\dot \theta \dot \phi\cot\theta+ \frac{GI\omega}{c^2}  \frac{\cot\theta}{r^3}\dot \theta-\frac{GI\omega}{2c^2}  \frac{\dot r}{r^4}, \label{sphe3}
\end{eqnarray}
which for $\theta=\pi/2$, are reduced to
\begin{eqnarray}
\ddot r =r\dot \phi^{2}
-\frac{GM}{r^2}+\frac{GI\omega}{2c^2}\frac{\dot \phi}{r^2}\label{B}, \\
\ddot \phi=-2\frac{\dot{r}}{r}\dot\phi  -\frac{GI\omega}{2c^2}  \frac{\dot r}{r^4}\label{A}.
\end{eqnarray}
\section{Equivalency Relation}
Due to condition (\ref{Sch}),  it is easily seen that the second terms in equations  $\left((\ref{eq:angular}), (\ref{A})\right)$ (right hand side) is negligible  compared to the first term\rlap.\footnote{Those terms are order of $r^{-4}$.} After neglecting this term, by comparing the third terms in equations (\ref{eq:angular1}), (\ref{B}), we get the equivalence relation
\begin{equation}\label{Equivalence1}
k\Theta=\frac{GI\omega}{2c^2},
\end{equation}
which by  $k=GmM$, reads
\begin{equation}\label{Equivalence}
\Theta=\frac{a^2\omega}{5mc^2}\propto \frac{a^2\omega}{mc^2}.
\end{equation}
Note that the equation (\ref{Equivalence1}) is not a relation  to determine the values of  its ingradients, say $\Theta$, as it appears in equation (\ref{Equivalence}). In fact, it means that, if the relation (\ref{Equivalence1}) is true, then the motion equations in the both space (NC and usual) are the same and equation (\ref{Equivalence})  was just to compare the values of its ingredients and specially to have an understanding of order of $\theta$.

As a last point, we note that, it is true that the equivalency relation (\ref{Equivalence1}) obtained based on ignoring the terms of order $r^{-4}$ in equations $(\ref{eq:angular})$ and $(\ref{A})$, but without actually removing this term, we would get the same result. This can be explained as follows:

We first note that the constant coefficients of the terms involving $\Theta$, in equations (\ref{eq:angular1}) and (\ref{eq:angular}),  differ by a factor of two. Also, since the NC parameter $\Theta$ is order of
$10^{-58} m^2$, \footnote{This gives observable corrections to the movement of the
solar system.}\cite{Romero} or  $10^{-62} m^2$ \cite{Mirza}, therefore, a factor of two in the NC parameter can not make a considerable effect on the final result. Of course, such an argument can be confirmed by the graphical representation of the particle trajectories which has been illustrated in  figure 1. In fact, the figure shows the motion paths (polar plots) corresponding to  two  set of motion equations:
\begin{center}(\ref{eq:angular1}), (\ref{eq:angular})\hspace{15mm}NC space\end{center}
and
\begin{center}(\ref{B}), (\ref{A})\hspace{15mm}GM space,\end{center}
\begin{figure}[h]
\begin{center}
\epsfig{figure=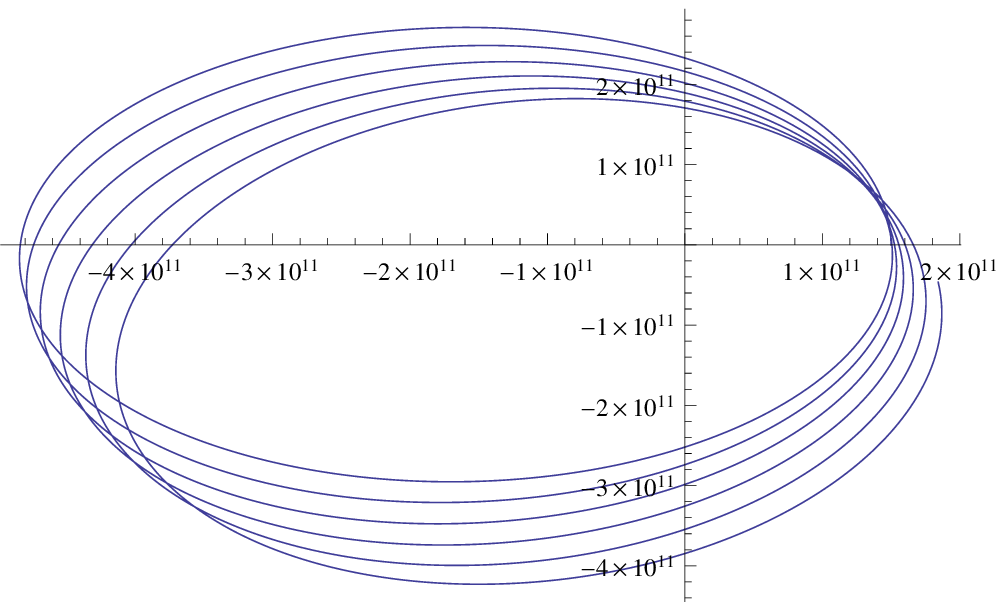,width=6cm}\hspace{.5cm}\epsfig{figure=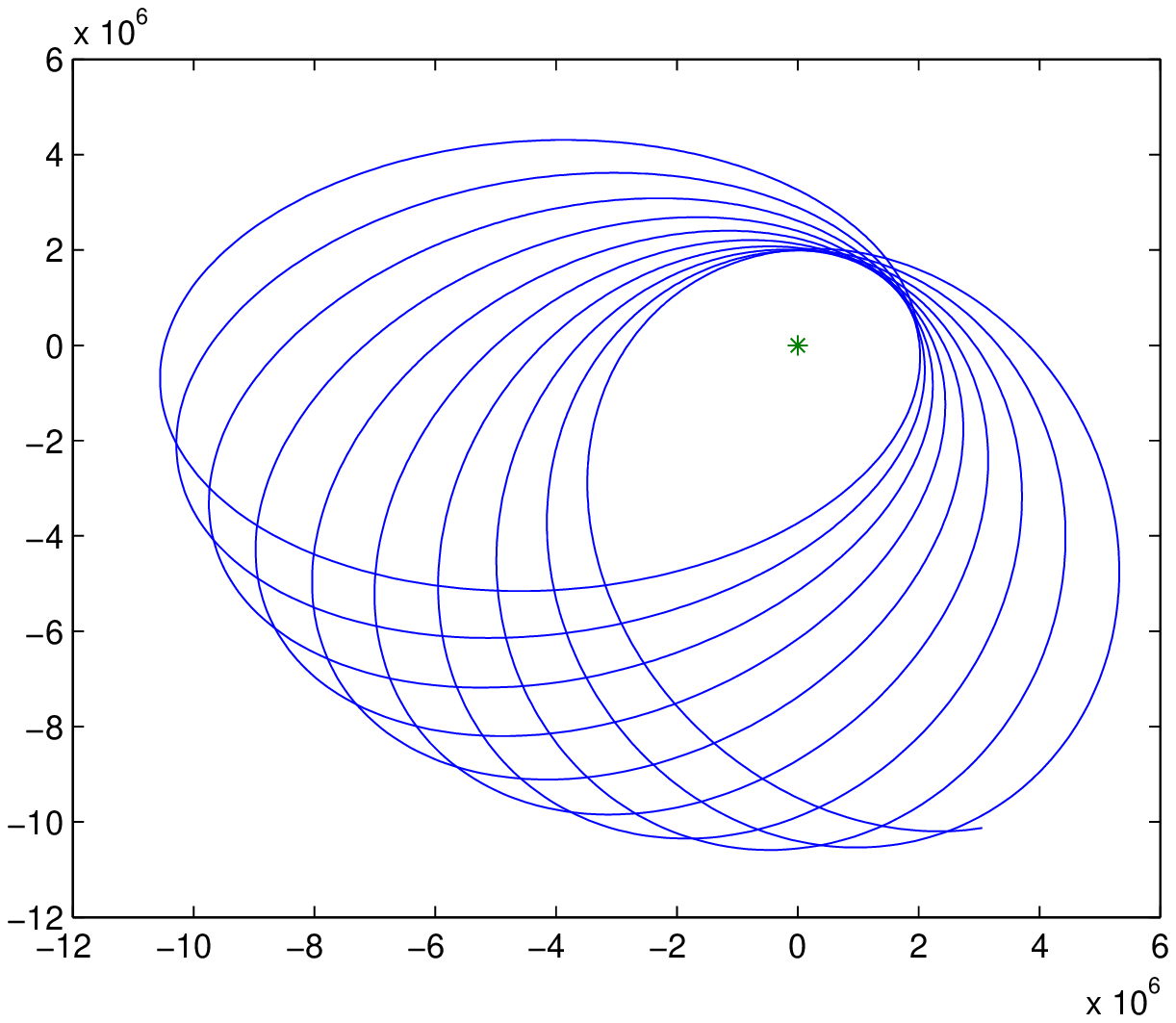,width=6cm}

{\footnotesize \textbf{Fig.1}: The trajectory of a test particle is influenced by
forces due to  the NC effects (left) and  the GM field (right). }
\end{center}
\end{figure}
without removing the second terms (in the second equations). As plots show the two paths are qualitatively analogous\rlap.\footnote{These plots represent the  counterclockwise (precession) motion of ellipses.}  Indeed, these plots illustrate the precession motion (the motion of perigee in orbital plan) arisen from the NC effects (Fig.1-left) and  intrinsic motion of the sphere (Fig.1-right).

The values of parameters for the plots in Fig.1 are  adapted to \emph{SI} unites, but for the right plot, we use the unit $c=1$. The values are as follows:

In the NC frame, the values of  mass for the central and orbiting  objects, are $M=2\times 10^{30}, m=6\times 10^{24}$ respectively and the NC parameter  value is $\theta=10^{-26}$. In the usual space the values  mass for the rotating sphere and orbiting particle, are   $M=1.36\times 10^{24}, m=1$ respectively. The radius  and angular velocity of the sphere are $ R=2\times 10^{4}, \omega=2\times 10^{-8}$  respectively. 
\section{conclusion}
Gravito--magnetism is an old subject in general relativity around which many well written papers
exist. However, the effects of accompanying non–commutativity lends some weight to its significance.
However, its use in the framework of gravito--magnetism has been somewhat
left on the side. It is thus interesting to study within this framework. Hence, in this work, we have
investigated the NC effects corresponding to the space sector of the phase space. The investigation
is relevant to the equivalence between the NC effects and the intrinsic GM field. For this purpose, the motion equations of a test particle in the usual space and NC space are compared with each other. In the usual space, the gravitational (gravito-electromagnetic) fields are produced by a slowly spinning sphere and in the NC space,  it is the responsibility of a static rigid sphere to produce the field. The comparison shows that, the effects of NC parameter  is similar to
the rotation of the sphere.  The main conclusion  can be summarized in
the short following statement:

\begin{center}\emph{The NC effects  mimic the intrinsic GM effects and vise versa}.\end{center}
%

%
\end{document}